\documentclass[12pt,preprint]{aastex}

\shorttitle{Molecular outflows in ULIRGs}
\shortauthors{Sturm et al.}
\usepackage{graphicx}
\usepackage{amsfonts}
\begin{document}
   \title{Massive molecular outflows and negative feedback in ULIRGs\\ observed by {\it Herschel}\thanks{{\it Herschel} is an ESA space observatory with science instruments provided by European-led
Principal Investigator consortia and with important participation from NASA.}-PACS}

\author{
E. Sturm\altaffilmark{1},
E. Gonz{\'a}lez-Alfonso\altaffilmark{2},
S. Veilleux\altaffilmark{3}
J. Fischer\altaffilmark{4},
J. Graci{\'a}-Carpio\altaffilmark{1},
S. Hailey-Dunsheath\altaffilmark{1},
A. Contursi\altaffilmark{1},
A. Poglitsch\altaffilmark{1},
A. Sternberg\altaffilmark{5},
R. Davies\altaffilmark{1},
R. Genzel\altaffilmark{1},
D. Lutz\altaffilmark{1},
L. Tacconi\altaffilmark{1}
A. Verma\altaffilmark{6},
R. Maiolino\altaffilmark{7},
J. A. de Jong\altaffilmark{1}
}

\altaffiltext{1}{Max-Planck-Institute for Extraterrestrial Physics (MPE), Giessenbachstra{\ss}e 1, 85748 Garching, Germany}
\altaffiltext{2}{Universidad de Alcal{\'a} de Henares, 28871 Alcal{\'a} de Henares, Madrid, Spain}
\altaffiltext{3}{Department of Astronomy, University of Maryland, College Park, MD 20742, USA}
\altaffiltext{4}{Naval Research Laboratory, Remote Sensing Division, 4555 Overlook Ave SW, Washington, DC 20375, USA}
\altaffiltext{5}{Tel Aviv University, Sackler School of Physics \&\ Astronomy, Ramat Aviv 69978, Israel}
\altaffiltext{6}{Oxford University, Dept. of Astrophysics, Oxford OX1 3RH, UK}
\altaffiltext{7}{INAF-Osservatorio astronomico di Roma, via Frascati 33, 00040 Monteporzio Catone, Italy}
\email{sturm@mpe.mpg.de}


\begin{abstract}
{Mass outflows driven by stars and active galactic nuclei are a key element in many current models of galaxy evolution. They may produce
the observed black hole-galaxy mass relation
and regulate and quench both star formation in the host galaxy and black hole accretion.
However, observational evidence of such feedback processes through outflows of the bulk of the star forming molecular gas
is still scarce. Here we report the detection of massive molecular outflows, traced by the hydroxyl molecule (OH), in far-infrared
spectra of ULIRGs obtained with {\it Herschel}-PACS as part of the SHINING key project. In some of these objects the (terminal) outflow velocities exceed 1000 km/s, and their outflow rates (up to $\sim$1200 M$_\odot$/yr) are several times larger than their star formation rates.
We compare the outflow
signatures in different types of ULIRGs and in starburst galaxies to address the issue of the energy source (AGN or starburst) of these
outflows. We report preliminary evidence that ULIRGs with a higher AGN luminosity (and higher AGN contribution to L$_{IR}$) have higher terminal
velocities
and shorter gas depletion time scales. The outflows in the observed ULIRGs are able to expel the cold gas reservoirs from the centres of these objects
within $\sim$10$^6$--10$^8$ years.
}
\end{abstract}


   \maketitle
%

\section{Introduction}

Gas-rich galaxy merging may trigger major starbursts, lead to the formation of elliptical galaxies, and account for the formation and growth of
supermassive black holes (BHs; e.g., Sanders et al. 1988; Hopkins et al. 2009). This merger-driven evolutionary scenario starts with a completely
obscured ultraluminous infrared galaxy (ULIRG).
As the system evolves, the obscuring
gas and dust
is gradually dispersed, giving rise
to dusty QSOs and finally to completely exposed QSOs.
Powerful winds, driven by the central quasar or the surrounding starburst, have been invoked to quench the growth of both the BH and
spheroidal component and explain the tight BH-spheroid mass relation (e.g., di Matteo et al. 2005, Murray et al. 2005). These winds are purported to inhibit star formation
in the merger remnants (``negative mechanical feedback''; e.g. Veilleux et al. 2005 for a review), and to create a population of red gas-poor
ellipticals, thereby explaining the bimodal color distribution observed in large galaxy surveys (e.g., Kauffmann et al. 2003).

Finding observational evidence of such feedback processes in action is one of the main challenges of current extragalactic
astronomy. While outflows have been observed frequently in many starbursts and QSOs, so far they have been detected mostly in the ionized and neutral atomic gas component.
To inhibit star formation in the host galaxy, outflows have to affect the {\it molecular} gas out of which stars form.
Few detections of molecular outflows have been reported so far (e.g. Baan et al. 1989, Walter et al. 2002, Sakamoto et al. 2009).
In this Letter we demonstrate that far-infrared molecular spectroscopy with Herschel-PACS of (ultra-) luminous infrared galaxies
is providing a breakthrough in identifying and analyzing massive molecular outflows. Our recent OH-absorption observations have revealed a $>$1000 km/s
molecular outflow in the closest quasar known, Mrk\,231 (Fischer et al. 2010). Independent, spatially resolved CO-emission observations of Mrk\,231
with the IRAM/PdB mm-wave interferometer (Feruglio et al. 2010) have confirmed this outflow with inferred mass outflow rates of $\sim$300--2200
M$_\odot$/yr, significantly larger than the current star formation rate (SFR$\sim$100\,M$_\odot$/yr) in the host galaxy. We now show that molecular outflows are indeed a common phenomenon in many of the luminous major mergers in our sample, reaching outflow velocities of 1000 km/s and outflow rates up to $\sim$1000 M$_\odot$/yr in some of them.

   \begin{figure*}
   \centering
   \includegraphics[width=16cm]{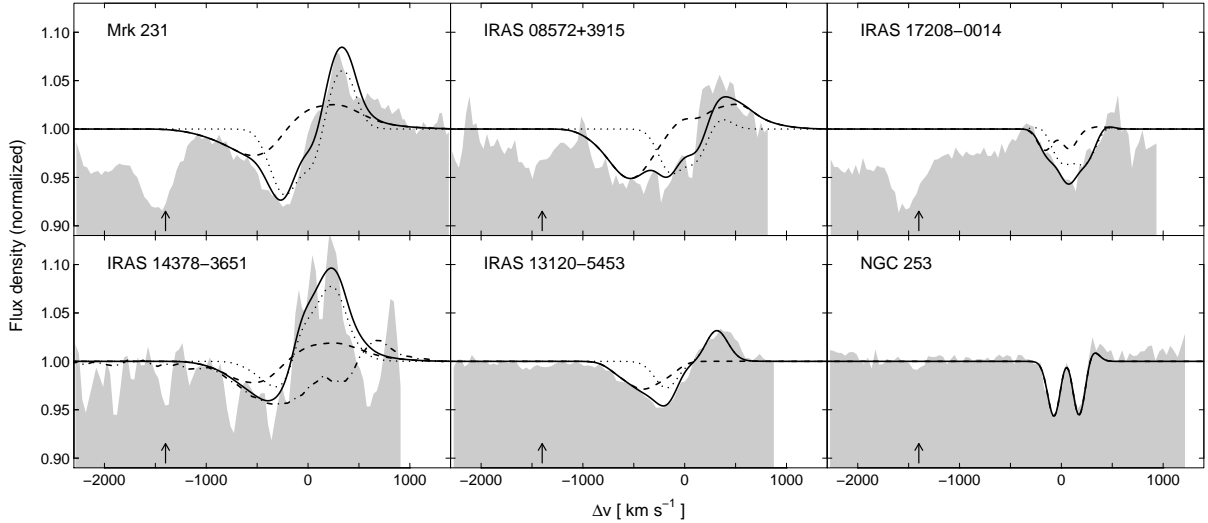}
   \caption{
   Observed PACS spectra (continuum normalized) of the OH transition at 79 $\mu$m ({\it grey}). Overplotted are the low velocity
   ({\it dotted}) and high velocity ({\it dashed}) fit components and the total fit ({\it full}). The arrow indicates the rest position of H$_2$O 4$_{23}$-3$_{12}$.
   The dash-dotted line for IRAS 14378 shows the observed spectrum of the OH transition at 119 $\mu$m for this object.}
              \label{Fig1}%
    \end{figure*}
%

\section{Observations and Data Reduction}

The data presented here are part of the {\it Herschel} guaranteed time key program SHINING, a study of the far-infrared properties of the ISM in
starbursts, Seyfert galaxies and infrared luminous galaxies.
Here we present velocity-resolved line profiles of the OH 79$\mu$m cross ladder, ground-state doublet for a small subset of our ULIRG sample and a
comparison starburst galaxy. The observations were obtained with the PACS far-infrared spectrometer (Poglitsch et al. 2010) on board Herschel (Pilbratt et al. 2010). Our data set also contains other OH transitions, but the 79$\mu$m line was observed first and is common to all objects, as it can be obtained simultaneously (in second order) with the [C\,II] 157$\mu$m line (in first order).  At 79$\mu$m the PACS
resolution is $\sim$140 km/s. The data reduction was done using the standard PACS reduction and calibration pipeline (ipipe) included in HIPE 5.0.
However, for the final calibration we normalized the spectra to the telescope flux (which dominates the total signal, except for NGC\,253) and
re-calibrated it with a reference telescope spectrum obtained from dedicated Neptune observations during the Herschel performance verification
phase.
All of our objects (except NGC\,253) are point sources for PACS. In the following we use the spectrum of the central 9$\arcsec$x9$\arcsec$ spatial
pixel (spaxel) only, applying the point source correction factors (PSF losses) as given in the PACS documentation. We have verified this approach
by comparing the resulting continuum flux density level to the continuum level of all 25 spaxels combined (which is free of PSF losses and pointing
uncertainties). In all cases the agreement is excellent, however the central spaxel alone provides better S/N.
We note for completeness that for NGC\,253 the total OH 79$\mu$m line profile summed over all 5$\times$5 spaxels yields emission, consistent with the ISO-LWS observations by Bradford et al. (1998).

In a next step we have performed a continuum (spline) fit. Due to the limited wavelength coverage these fits are somewhat subjective. To help define continuum points and potential additional spectral features (such as the
H$_2$O absorption line at 78.74$\mu$m, indicated with an arrow in Fig. \ref{Fig1}) we have
used our full range spectra of Arp\,220 and NGC\,4418. These two sources will be analyzed in detail in forthcoming papers, but preliminary data points for Arp\,220 are included in Figures 2 and 3. We note here that NGC\,4418 shows signatures of an inflow.

\begin{table*}
\caption{Target properties, outflow rates and outflow velocities (1$\sigma$ uncertainties in brackets)}             
\label{Tab1}      
\centering                          
\begin{tabular}{l | c | c | c | c| c| r| r| r}
\hline\hline                 
\rule[0mm]{0cm}{5mm} Source   &  SFR       & $\alpha^{\mathrm{a}}$ & L$_{AGN}$ & $M_{gas}^{\mathrm{b}}$  &  $\dot{M}^{\mathrm{c}}$ & v$_{peak}^{\mathrm{d}}$ & v$_{85\%}^{\mathrm{e}}$ & v$_{max}^{\mathrm{f}}$  \\    
                  & M$_{\odot}/yr$ & \% & 10$^{11}$L$_\odot$ & 10$^{9}$M$_{\odot}$ & M$_{\odot}$/yr & km/s & km/s  &  km/s  \\
\hline                        
\rule[0mm]{0cm}{4mm}Mrk\,231          &  101 (15)  & 71 (11) & 28 (4)    & 4.2 (1.3)  & 1190$^{+4700}_{-890}$ & -600  & -660 &-1170 \\
\rule[0mm]{0cm}{4mm}IRAS\,08572+3915  &   42 (6)   & 72 (11) & 12 (2)    & 1.3 (0.4)  &  970$^{+2900}_{-730}$ & -700  & -740 &-1260 \\
\rule[0mm]{0cm}{4mm}IRAS\,13120-5453  &  168 (25)  &  9 (1.4)& 1.8 (0.3) & 5.8 (1.7)  &  130$^{+390}_{-95}$   & -520  & -600 &-860 \\
\rule[0mm]{0cm}{4mm}IRAS\,14378-3651  &  $>$79     & $<$45   & $<$7.2    & 4.2 (1.3)  &  740$^{+2200}_{-550}$ & -800  & -860 & -1170 \\
\rule[0mm]{0cm}{4mm}IRAS\,17208-0014  &  274 (41)  & 11 (1.7)& 3.4 (0.5) & 12.2 (3.7) &   90$^{+270}_{-65}$   & -100  & -170 & -370 \\
\rule[0mm]{0cm}{4mm}NGC\,253          &  1.7 (0.3) &  0      & 0         & 0.7 (0.2)  &  1.6$^{+4.8}_{-1.2}$  &  -75  & -130 & -280 \\
\hline
\end{tabular}

\begin{list}{}{}
\item[ ] {\footnotesize $^{\mathrm{a}} $ Fraction of the AGN contribution to L$_{bol}$, where L$_{bol}$ = 1.15$\times$L$_{IR}$;
$^{\mathrm{b}} $ Gas mass (taken from Graci\'a-Carpio et al. 2011);
$^{\mathrm{c}} $ mass outflow rate (see footnote of Table \ref{Tab2});
$^{\mathrm{d}} $ peak velocity of the blue shifted high velocity component (relative to systemic velocities);
$^{\mathrm{e}} $ velocity for which 85\% of the outflowing gas has lower (absolute) velocities;
$^{\mathrm{f}} $ Terminal velocity. Estimated uncertainty for all velocities: $\pm$150km/s.}

\end{list}
\end{table*}

\section{Targets}
\label{sect:targets}

For this first study of outflow signatures in our data we use a sub-sample that is mainly constrained by the observing schedule of
Herschel, but that covers a broad range of AGN and starburst activity, including a starburst template (NGC\,253), a cold, starburst dominated ULIRG
(IRAS\,17208-0014,) warm ULIRGs (S25/S60$>$0.1) and/or ULIRGs with strong AGN contributions (Mrk\,231, IRAS\,13120-5453, IRAS\,14378-3651), and a
heavily obscured ULIRG (IRAS\,08572+3915), which hosts a powerful AGN (e.g. Veilleux et al. 2009, V09).

\section{Results and Discussion}

Figure \ref{Fig1} shows the (continuum normalized) OH 79$\mu$m line spectra for all objects. For NGC\,253 we show the central spatial pixel only.
The Mrk\,231 spectrum is taken from Fischer et al. (2010) and is repeated here for completeness. In all cases we detect P-Cygni profiles typical of
outflows, with blue shifted absorption and red shifted emission features, of more than 1000 km/s in some cases.
For comparison with the literature we list in Table \ref{Tab1} various measures of outflow velocities (relative to the system velocity of the blue component of the OH doublet): the peak velocity (v$_{peak}$), the maximum (terminal) velocity (v$_{max}$), and v$_{85\%}$ for which 85\% of the outflowing gas has lower (absolute) velocities. The uncertainties of these velocities are dominated by the uncertainties in the continuum definition, the S/N ratio in the spectra, and the spectral resolution. We estimate an overall error of $\pm$150\,km/s. The S/N in the IRAS\,14378-3651 spectrum is relatively low, but the high terminal velocity is confirmed by the OH 119 $\mu$m transition (see Figure \ref{Fig1}).

These high OH outflow velocities may be the long-sought conclusive evidence of powerful mechanical feedback from vigorous star formation and/or accreting central black holes.
The possible feedback and outflow mechanisms (e.g. winds from SNe, radiation pressure) are debated in the literature (see, e.g., the review by Veilleux et al. 2005).
It is not clear, if such mechanisms could indeed be sufficient to power outflows
that are strong enough to significantly affect the host galaxy and to actually quench the star formation in these objects. It is also unclear from
the models whether it is possible to distinguish AGN driven outflows from stellar-driven outflows observationally (see, e.g., Hopkins \& Elvis 2010). In
the following we adopt an empirical approach with our new data.

\subsection{Are the strong outflows we observe driven by the AGN rather than by the star formation in these objects?
}
Rupke et al. (2005a,b,c) and Krug, Rupke \& Veilleux (2010)
have studied large samples of AGN and star forming galaxies in neutral gas (blueshifted optical Na ID 5890, 5896 {\AA} absorption features). They
found that, for fixed SFR, ULIRGs with higher AGN fractions have higher neutral gas outflow velocitites,
reaching velocities well above 1000 km/s in some broad-line AGN (see also, e.g., Heckman et al. 2000, Martin 2005, 2006, Thacker et al. 2006).
Theoretical models predict that SNe driven outflows cannot reach velocities higher than 500-600 km/s (e.g. Martin 2005, Thacker et all. 2006).
Predictions of v$_{max}$ from models of outflows driven by radiation pressure from a starburst (or AGN) are, however, less certain.
The terminal velocity we measure in the OH outflow of our starburst template NGC\,253 is $\sim$300
km/s. IRAS\,17208-0014, a starburst dominated ULIRG with little AGN contribution has only a slightly higher
v$_{max}$\,(370 km/s). In significant contrast to this the terminal velocities of the OH outflow in the two AGN-dominated ULIRGs (Mrk\,231 and
IRAS\,08572+3915) are well above 1000 km/s.
Thus the OH outflow velocity could be a very promising tool to distinguish AGN driven outflows from starburst driven outflows, with
AGN-dominated outflows reaching much higher velocities.

\begin{figure}
   \centering
   \includegraphics[width=14cm]{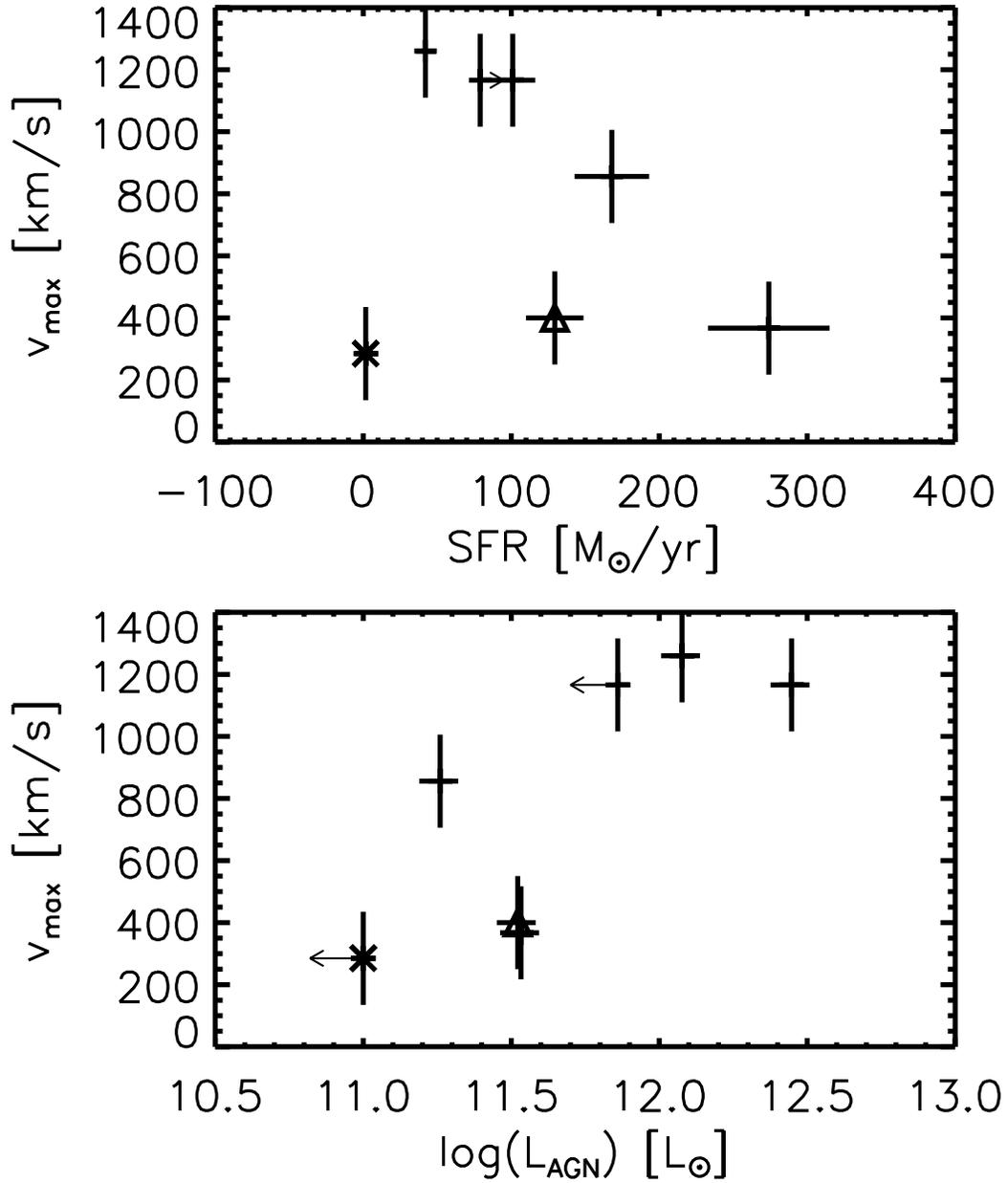}
   \caption{
   Maximum outflow velocities (terminal velocities) as function of star formation rate (upper panel), and AGN luminosity (lower panel).
   The asterisk denotes NGC\,253, the triangle Arp\,220.  }
              \label{Fig2}%
    \end{figure}

In Figure \ref{Fig2} we compare the terminal outflow velocities to the star formation rates (SFRs) and AGN-luminosities
of our objects. SFRs are calculated from the IR luminosities
(based on the calibration of Kennicutt 1998, but using a Chabrier IMF rather than a Salpeter IMF),
i.e. SFR = (1-$\alpha)\times$10$^{-10}$L$_{IR}$, applying
AGN correction factors $\alpha$ (the fraction of the contribution from the AGN to L$_{bol}$, where L$_{bol}$ = 1.15$\times$L$_{IR}$ as in V09).
These factors are individually derived from Spitzer mid-IR diagnostics provided by the QUEST programme (V09). For IRAS\,13120-5453 and
IRAS\,14378-3651 (which are not part of V09) we calculated AGN fractions using method 1 from V09, based on the observed [O\,IV]/[Ne\,II] ratio (and
the observed upper limit for IRAS\,14378-3651) from Farrah et al. (2007). AGN luminosities are then calculated as L$_{AGN}$ =
$\alpha\times$L$_{bol}$. AGN luminosities and SFRs are listed in Table \ref{Tab1}.

For star formation driven outflows one might expect the outflow velocities to scale with the star formation rates (e.g. Tremonti et al. 2007). We do not see such
a correlation (Figure \ref{Fig2}, upper panel). Instead, we see a rough correlation of $v_{max}$ with L$_{AGN}$ (Figure \ref{Fig2}, lower panel), consistent with the idea of the
high velocity outflows being powered mostly by (radiation pressure from) the AGN.

Very energetic outflows have been found in radio galaxies at high $z$ (e.g. McCarthy et al. 1996, Best et al. 2000,  Nesvadba et al. 2009). Radio jets could in principle be
driving the outflows we see. However, nearly all of the objects presented here
are radio quiet, so this energy source can safely be assumed to be negligible. Only in Mrk\,231 does a radio jet contribute to the Na\,{\small I}\,D outflow (Rupke \& Veilleux 2011).

\subsection{Does the outflow carry sufficient molecular gas to remove
the star formation fuel and actualy quench the star formation?
}

\begin{table*}
\caption{(Preliminary) model fit results (errors are discussed in the text)}             
\label{Tab2}      
\centering                          
\begin{tabular}{l | c | c | c | c| c| c}
\hline\hline                 
\rule[0mm]{0cm}{4mm} Source           &component& R$_{in}^{\mathrm{c}}$ & n(OH)$_{in}^{\mathrm{d}}$ &  $f^{\mathrm{e}}$ &
$\theta^{\mathrm{f}}$  & $<$N(OH)$>^{\mathrm{g}}$ \\
                                      &     &    pc   & 10$^{-4}$cm$^{-3}$    &       &   degrees &    10$^{16}$cm$^{-2}$  \\
\hline                        
\rule[0mm]{0cm}{4mm}Mrk\,231          & HVC$^{\mathrm{a}}$ &   105 &  7    &  0.6 &     90  &    3 \\
                                      & LVC$^{\mathrm{b}}$ &   115 &  5    &  1.0 &     61  &    5 \\
\rule[0mm]{0cm}{4mm}IRAS\,08572+3915  & HVC &   110 &  20    &  0.2 &     77  &    4 \\
                                      & LVC &   110 &  40    &  0.3 &     69  &    4 \\
\rule[0mm]{0cm}{4mm}IRAS\,13120-5453  & HVC &   210 &  9    &  0.3 &     43  &    3 \\
                                      & LVC &   210 &  1    &  0.5 &     90  &    2 \\
\rule[0mm]{0cm}{4mm}IRAS\,14378-3651  & HVC &   100 &  3    &  0.5 &     90  &    1 \\
                                      & LVC &   120 &  2    &  1.0 &     90  &    1 \\
\rule[0mm]{0cm}{4mm}IRAS\,17208-0014  & LVC &   110 &  6    &  1.0 &     57  &    2 \\
\rule[0mm]{0cm}{4mm}NGC\,253          & LVC &    90 &  0.5    &  1.0 &     51  &    1 \\
\hline
\end{tabular}

\begin{list}{}{}
\item[ ] {\footnotesize
$^{\mathrm{a}} $ high velocity component;
$^{\mathrm{b}} $ low velocity component;
$^{\mathrm{c}} $ inner shell radius;
$^{\mathrm{d}} $ OH density at R$_{in}$;
$^{\mathrm{e}} $ filling factor;
$^{\mathrm{f}} $ half opening angle (i.e. $\theta=90^{\circ}$ for a full 4$\pi$ coverage);
$^{\mathrm{g}} $ OH column density N(OH)$\times f$. \\
These values are used to compute the mass outflow rate in Table \ref{Tab1}:  \\
$\dot{M}\sim M_{gas}/t_{dyn} \sim 4\pi\times n(OH)_{in}/\chi (OH)\times m_{H_2}\times R_{in}^2\times f\times g\times v$, where $t_{dyn}\sim R/v$, $g$ is a function of the opening angle $\theta$ and
$\chi (OH) = 5\times 10^{-6}$ is the OH abundance relative to H$_2$. Outflow rates in Table \ref{Tab1} are the sum of the components (HVC+LVC).
A more detailed description of a refined model will be given in a future paper.}

\end{list}
\end{table*}

To compute mass outflow rates ($\dot{M}$) for comparison with SFRs
we have modeled the observed spectra using the radiative transfer code
described in Gonz\'alez-Alfonso \& Cernicharo (1999).
The outflow is modeled as concentric expanding shells around a nuclear
continuum source, allowing for each source one to three
components with different velocity gradients and distances to the
central source. Such geometry reproduces quite naturally the red-shifted emission (which is produced in the receding cocoons),
and the blue-shifted absorption (from the approaching parts).

The density profile for each velocity component is determined through mass
conservation assuming a stationary outflow. With the exception of
NGC\,253, where we fit the nuclear continuum emission from the observed PACS
spectrum, the nuclear FIR in all other sources is modeled
as the warm component fitted in Mrk\,231 (Gonz\'alez-Alfonso et al. 2010).
Free parameters are the inner and outer radii and velocity
field of each velocity component, the OH density at the inner radius,
the covering factor of the continuum FIR source, and the solid angle of the
outflow.
Because the OH energy levels are radiatively pumped in the outflows,
transitions at different wavelengths and energy levels, in combination
with continuum component fits, yield crucial information about the
radial location where the lines are formed. Besides the 79 $\mu$m
doublet, the 119
(also ground-state) and 65 $\mu$m
($E_{\mathrm{lower}}=290$ K) doublets were also
observed in all our sources, except for IRAS\,17208.
The excitation of the high-lying 65 $\mu$m doublet requires high
far-IR radiation densities, indicating that the line is tracing the inner,
highly excited, molecular region of the outflowing material, close to the
nuclear source (Gonz\'alez-Alfonso et al. 2008). Models that account for the 65 $\mu$m line
absorption simultaneously reproduce the two ground-state lines as well, indicating
that the three observed OH lines - in contrast to the atomic Na\,ID lines, which are extended on
galactic scales in some cases - are mostly sensitive to the inner
$\sim0.5$ kpc of the outflow. For IRAS\,17208 only the 79
$\mu$m line is currently available, but we have assumed a similarly compact outflow in OH.
The OH columns are typically several $10^{16}$ cm$^{-2}$.
In Table \ref{Tab2} we list the fit results. Figure \ref{Fig1} shows the various components of the spectral fits
on top of the observed spectra.

We derive
mass-loss rates $\dot{M}$ (see Table \ref{Tab1} and footnote of Table \ref{Tab2}) from the OH density profile of the outflowing gas,
the gas velocity in each component, and the adopted OH abundance.
The resulting mass-loss rates are higher than 700 M$_{\odot}$/yr in
those sources with high $v_{max}$. The uncertainty
in $\dot{M}$ is dominated by the adopted OH abundance,
the properties of the underlying FIR continuum source, and the outflow
geometry. We adopt an overall uncertainty of 4, derived from a study of the dependency
of the fit results on the  various model assumptions. The OH abundance relative to H$_2$ is assumed to be
$5\times10^{-6}$, based on modeling of multi-transition OH
observations of the Galactic giant molecular cloud Sgr\,B2 (Goicoechea \& Cernicharo 2002).
While OH abundances could reach values significantly lower in different
environments (e.g. in some of the models of Sternberg \& Dalgarno 1995 and
Meijerink \& Spaans 2005, or the observations by Watson et al. 1985), we
expect a relatively high OH abundance close to the nuclear region
(Gonz\'alez-Alfonso et al. 2008) and note that $\dot{M}$ increases with decreasing OH
abundance. On the other hand, even a high OH abundance of $10^{-5}$ leads
to mass loss rates of several hundred M$_{\odot}$/yr. Further details on the
modeling will be given in a forthcoming paper.

%

   \begin{figure}
   \centering
   \includegraphics[width=14cm]{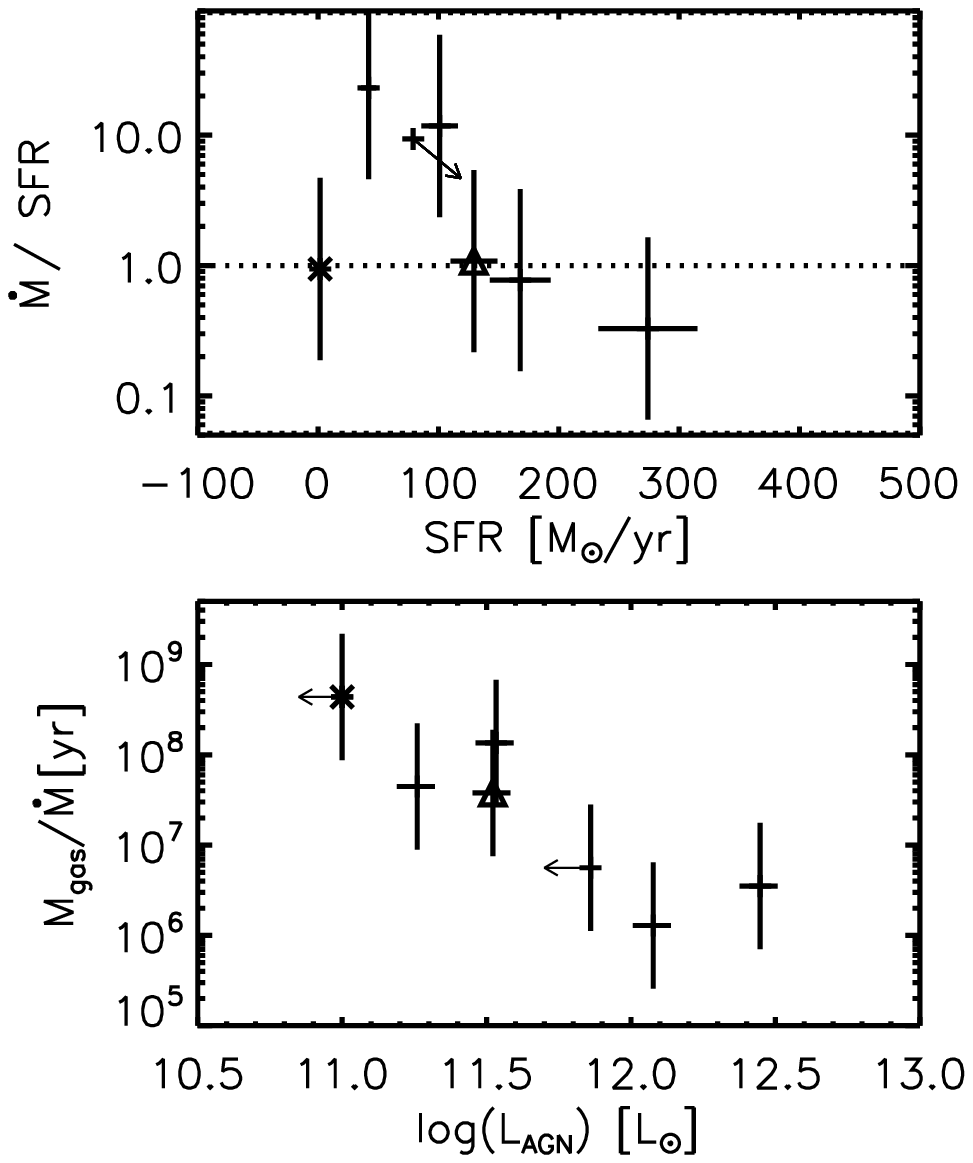}
   \caption{{\it upper panel}: the ratio of the mass outflow rate to the star formation rate versus SFR;
   {\it lower panel}: depletion time scale versus AGN luminosity. Symbols as in Figure \ref{Fig2}.
   }
              \label{Fig3}
    \end{figure}

The starburst-dominated objects of
our sample (NGC\,253, IRAS\,13120-5453, and IRAS 17208-0014) have values of $\dot{M}$ similar to their star formation rates. In contrast to this, the AGN-dominated
and/or warm ULIRGs, as well as the heavily obscured ULIRG, have outflow rates that are factors $\sim$4--20 larger than their SFR (see Table \ref{Tab1}
and Figure \ref{Fig3}, upper panel). The outflows we observe in the ULIRGs - if continued at the current rate - are able to completely expel the cold gas reservoirs from the centres of these objects
within $\sim$10$^6$--10$^8$ years (Figure \ref{Fig3}, lower panel). The gas depletion time is roughly inversely correlated with the AGN luminosity (and AGN fraction). This is consistent with the merger scenario where the highest outflow rates
are a short-lived, late, AGN-dominated stage in the merger evolution. We note, however, that the lowest time scales we find (few$\times$10$^6$
years) appear to be inconsistent with star formation/galaxy evolution models. We may have underestimated the time scales: a higher gas mass by a factor 2 and a higher OH abundance by a factor 2 would result in 4 times longer time scales (which would still be within the error bars of figure 3). A fundamental change in the basic model assumptions, like the geometry of the outflow, could also change the outflow rates.
Our current models are consistent with wide opening angles for the
outflowing gas, but this is hard to prove with PACS spectra alone (see below). For Mrk231, however, the high outflow rate is consistent (within a factor 2) with the independent CO observations (PdBI mm-interferometry) by Feruglio et al. 2010. Very recently the high rates and the wide angle of the outflow have also been confirmed by optical spectroscopy through the analysis of broad wings of H$\alpha$ and Na\,{\small I}\,D
absorption (Rupke \& Veilleux 2011).
Finally, the outflow may not continue at such a high rate, since - as gas and dust are removed -
the outflows may be quenched as the radiation pressure decreases. Therefore, our depletion time scales are
time {\it scales} rather then absolute durations.

We do not have the spatial resolution with Herschel-PACS to study the spatial distribution of the outflowing material directly
(although observations of additional OH transitions with higher energy level, e.g. at 84 and 163 $\mu$m, could help to further
constrain the geometry of the models).
True spatial information on the molecular component can only come from interferometric mm-observations (e.g. Feruglio et al. 2010). While the
observed spectra shown here clearly exhibit high velocities, powerful molecular outflows, independent of our modelling, the conclusions based on
outflow rates and AGN fractions need further confirmation from increased statistics. Our ongoing observations will cover a much larger sample of both starburst and AGN dominated (U)LIRGs than shown here. Combined with mm-interferometric follow up observations we will be able to better constrain our OH outflow models and further investigate the surprisingly low depletion time scales for some of the objects.
The final sample should then allow us to study potential trends in the outflow characteristics. Examples of such trends could be a different or tighter correlation of the outflow velocity with AGN luminosity than with star formation rate, or the extent to which AGN fractions and outflow velocities reflect different merger stages with evolving outflows creating lower and lower covering factors to the AGN.

\acknowledgements
We thank Dave Rupke for helpful discussions. Basic research in IR astronomy
at NRL is funded by the US ONR; J.F. also acknowledges support from the NHSC. E.G-A is a Research Associate at the Harvard-Smithsonian Center for Astrophysics. A.S. thanks the DFG for support via German-Israeli Project Cooperation grant STE1869/1-1.GE625/15-1.
PACS has been developed by a consortium of institutes led by MPE (Germany) and including UVIE (Austria); KU Leuven, CSL, IMEC (Belgium); CEA, LAM
(France); MPIA (Germany); INAF-IFSI/OAA/OAP/OAT, LENS, SISSA (Italy); IAC (Spain). This development has been supported by the funding agencies
BMVIT (Austria), ESA-PRODEX (Belgium), CEA/CNES (France), DLR (Germany), ASI/INAF (Italy), and CICYT/MCYT (Spain).



\begin{thebibliography}{}

\bibitem[Baan et al.(1989)]{1989ApJ...346..680B} Baan, W.~A., Haschick, A.~D., \& Henkel, C.\ 1989, \apj, 346, 680

\bibitem[Best et al.(2000)]{2000MNRAS.311...23B} Best, P.~N., R{\"o}ttgering, H.~J.~A., \& Longair, M.~S.\ 2000, \mnras, 311, 23

\bibitem[Bradford et al.(1999)]{1999ESASP.427..861B} Bradford, C.~M., et al.\ 1999, The Universe as Seen by ISO, 427, 861

\bibitem[Di Matteo et al.(2005)]{2005Natur.433..604D} Di Matteo, T., Springel, V., \& Hernquist, L.\ 2005, \nat, 433, 604

\bibitem[Farrah et al.(2007)]{2007ApJ...667..149F} Farrah, D., et al.\ 2007, \apj, 667, 149

\bibitem[Feruglio et al.(2010)]{2010A&A...518L.155F}
Feruglio, C., Maiolino, R., Piconcelli, E., Menci, N., Aussel, H., Lamastra, A., \& Fiore, F.\ 2010, \aap, 518, L155

\bibitem[Fischer et al.(2010)]{2010A&A...518L..41F} Fischer, J., et al.\ 2010, \aap, 518, L41

\bibitem[Goicoechea \& Cernicharo(2002)]{2002ApJ...576L..77G} Goicoechea, J.~R., \& Cernicharo, J.\ 2002, \apjl, 576, L77

\bibitem[Gonz\'alez-Alfonso \& Cernicharo(1999)]{gon99} Gonz\'alez-Alfonso, E., \& Cernicharo, J. 1999, ApJ, 525, 845

\bibitem[Gonz\'alez-Alfonso et al.(2008)]{gon08} Gonz\'alez-Alfonso, E., et al. 2008, ApJ, 675, 303

\bibitem[Gonz\'alez-Alfonso et al.(2010)]{gon10} Gonz\'alez-Alfonso, E., et al. 2010, AA, 518, L43

\bibitem[Graci\'a-Carpio et al. (2011)]{} Graci\'a-Carpio, J. et al. 2011, \apjl, 728, L7

\bibitem[Heckman et al.(2000)]{2000ApJS..129..493H} Heckman, T.~M., Lehnert, M.~D., Strickland, D.~K., \& Armus, L.\ 2000, \apjs, 129, 493

\bibitem[Hopkins et al.(2009)]{2009MNRAS.398..303H} Hopkins, P.~F., Murray, N., \& Thompson, T.~A.\ 2009, \mnras, 398, 303

\bibitem[Hopkins \& Elvis(2010)]{2010MNRAS.401....7H} Hopkins, P.~F., \& Elvis, M.\ 2010, \mnras, 401, 7

\bibitem[Kauffmann et al.(2003)]{2003MNRAS.341...54K} Kauffmann, G., et al.\ 2003, \mnras, 341, 54

\bibitem[Kennicutt(1998)]{1998ApJ...498..541K} Kennicutt, R.~C., Jr.\ 1998, \apj, 498, 541

\bibitem[Krug et al.(2010)]{2010ApJ...708.1145K} Krug, H.~B., Rupke, D.~S.~N., \& Veilleux, S.\ 2010, \apj, 708, 1145

\bibitem[Martin(2005)]{2005ApJ...621..227M} Martin, C.~L.\ 2005, \apj, 621, 227

\bibitem[Martin(2006)]{2006ApJ...647..222M} Martin, C.~L.\ 2006, \apj, 647, 222

\bibitem[McCarthy et al.(1996)]{1996ApJS..106..281M} McCarthy, P.~J., Baum, S.~A., \& Spinrad, H.\ 1996, \apjs, 106, 281

\bibitem[Meijerink \& Spaans(2005)]{2005A&A...436..397M} Meijerink, R., \& Spaans, M.\ 2005, \aap, 436, 397

\bibitem[Murray et al.(2005)]{2005ApJ...618..569M} Murray, N., Quataert, E., \& Thompson, T.~A.\ 2005, \apj, 618, 569

\bibitem[Nesvadba(2009)]{2009arXiv0906.2900N} Nesvadba, N.~P.~H.\ 2009, arXiv:0906.2900

\bibitem[Pilbratt et al.(2010)]{2010A&A...518L...1P} Pilbratt, G.~L., et al.\ 2010, \aap, 518, L1

\bibitem[Poglitsch et al.(2010)]{2010A&A...518L...2P} Poglitsch, A., et al.\ 2010, \aap, 518, L2

\bibitem[Rupke et al.(2005)]{2005ApJS..160...87R} Rupke, D.~S., Veilleux, S., \& Sanders, D.~B.\ 2005a, \apjs, 160, 87

\bibitem[Rupke et al.(2005)]{2005ApJS..160..115R} Rupke, D.~S., Veilleux, S., \& Sanders, D.~B.\ 2005b, \apjs, 160, 115

\bibitem[Rupke et al.(2005)]{2005ApJ...632..751R} Rupke, D.~S., Veilleux, S., \& Sanders, D.~B.\ 2005c, \apj, 632, 751

\bibitem[Rupke \& Veilleux(2011)]{2011ApJ...729L..27R} Rupke, D.~S.~N., \& Veilleux, S.\ 2011, \apjl, 729, L27

\bibitem[Sakamoto et al.(2009)]{2009ApJ...700L.104S} Sakamoto, K., et al.\ 2009, \apjl, 700, L104

\bibitem[Sanders et al.(1988)]{1988ApJ...325...74S} Sanders, D.~B., Soifer,
B.~T., Elias, J.~H., Madore, B.~F., Matthews, K., Neugebauer, G., \& Scoville, N.~Z.\ 1988, \apj, 325, 74

\bibitem[Sternberg \& Dalgarno(1995)]{1995ApJS...99..565S} Sternberg, A., \& Dalgarno, A.\ 1995, \apjs, 99, 565

\bibitem[Thacker et al.(2006)]{2006ApJ...653...86T} Thacker, R.~J., Scannapieco, E., \& Couchman, H.~M.~P.\ 2006, \apj, 653, 86

\bibitem[Tremonti et al.(2007)]{2007ApJ...663L..77T} Tremonti, C.~A., Moustakas, J., \& Diamond-Stanic, A.~M.\ 2007, \apjl, 663, L77

\bibitem[Veilleux et al.(2005)]{2005ARA&A..43..769V} Veilleux, S., Cecil, G., \& Bland-Hawthorn, J.\ 2005, \araa, 43, 769

\bibitem[Veilleux et al.(2009)]{2009ApJS..182..628V} Veilleux, S., et al.\ 2009 (V09), \apjs, 182, 628

\bibitem[Walter et al.(2002)]{2002ApJ...580L..21W} Walter, F., Weiss, A., \& Scoville, N.\ 2002, \apjl, 580, L21

\bibitem[Watson et al.(1985)]{1985ApJ...298..316W} Watson, D.~M., Genzel, R., Townes, C.~H., \& Storey, J.~W.~V.\ 1985, \apj, 298, 316



\end{thebibliography}
\end{document}